# Scanning probe microscopy with chemical contrast by nanoscale nuclear magnetic resonance.

T. Häberle[1], D. Schmid-Lorch[1], F. Reinhard[1] and J. Wrachtrup[1,2]

[1]3. Physikalisches Institut, Universität Stuttgart, Stuttgart, Germany

[2]Max Planck Institute for Solid State Research, Stuttgart, Germany

Scanning probe microscopy is one of the most versatile windows into the nanoworld, providing imaging access to a variety of sample properties, depending on the probe employed. Tunneling probes map electronic properties of samples[1], magnetic and photonic probes image their magnetic and dielectric structure[2,3] while sharp tips probe mechanical properties like surface topography, friction or stiffness[4]. Most of these observables, however, are accessible only under limited circumstances. For instance, electronic properties are measurable only on conducting samples while atomic-resolution force microscopy requires careful preparation of samples in ultrahigh vacuum[5,6] or liquid environments[7].

Here we demonstrate a scanning probe imaging method that extends the range of accessible quantities to label-free imaging of chemical species operating on arbitrary samples - including insulating materials - under ambient conditions. Moreover, it provides three-dimensional depth information, thus revealing subsurface features. We achieve these results by recording nuclear magnetic resonance signals from a sample surface with a recently introduced scanning probe, a single nitrogen-vacancy center in diamond. We demonstrate NMR imaging with 10 nm resolution and achieve chemically specific contrast by separating fluorine from hydrogen rich regions.

Our result opens the door to scanning probe imaging of the chemical composition and atomic structure of arbitrary samples. A method with these abilities will find widespread application in material science even on biological specimens down to the level of single macromolecules.

The development of a scanning probe sensor able to image nuclear spins has been a long and outstanding goal of nanoscience, proposed shortly after the invention of scanning probe microscopy itself[8]. To date, this goal is most closely met by magnetic resonance force microscopy (MRFM), an extension of atomic-force-microscopy with sensitivity to spins, which has successfully imaged nanoscale distributions of nuclear spins in three dimensions[9]. However, its operation is experimentally challenging, requiring low (sub-Kelvin) temperature and long (weeks/image) acquisition times, which has so far precluded its adoption as a routine technique. To surmount these problems, single electron spins with optical readout capability have been proposed as an alternative local probe for spin distributions[10]. This complementary approach has become a realistic prospect since recent research has established the nitrogen-vacancy center, a color defect in diamond[11], as a candidate system for this scheme[12,13]. This center serves as an atomic-sized magnetic field sensor, which has proven sufficiently sensitive to detect the field of single nuclear spins in its diamond lattice environment[14–16] as well as ensembles of $10$-$10^4$ spins in a nanometer-sized sample volume on the diamond surface[17–19]. Here we employ a single NV center as a scanning probe to image distributions of nuclear spins in an external sample. All our measurements are performed in the geometry of Fig. 1. At the heart of the experiment, a single NV center embedded approximately 5 nm below the surface of a bulk diamond serves as a nanoscale sensor to record the nuclear spin density in a sensing volume extending few nm above the diamond surface. To perform magnetic resonance imaging, a microscale (~µm sized) sample of NMR-active material is attached to the cantilever of a commercial scanning probe microscope and scanned across this single-pixel detection volume in contact mode. This technique effectively slides the NV sensor over the surface of the sample and is therefore conceptually equivalent to a scanning NV center attached to a probe tip, as it is commonly employed for imaging of static magnetic fields[12,20,21].

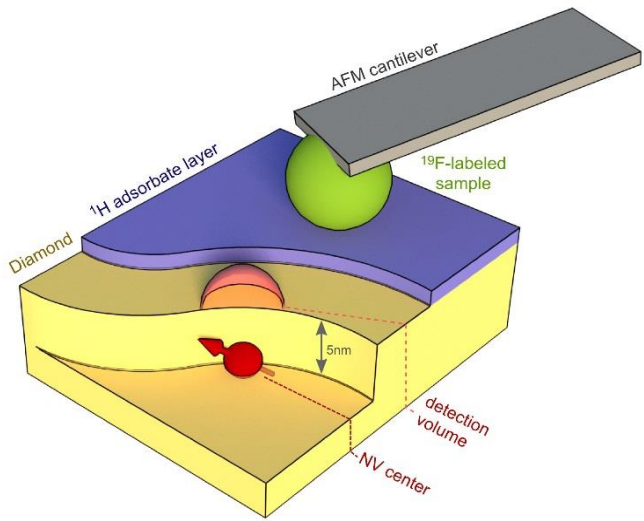

Fig 1 - **Experimental configuration**, A single NV center embedded few nm below the diamond surface (red spin) detects nuclear spins in a nanoscale detection volume above the diamond surface (red half-sphere). Imaging is performed by scanning a sample of NMR-active nuclear spins through this detection volume. In all experiments, a layer of proton rich organic adsorbates is present on the diamond surface, as it is unavoidable under ambient conditions.

To unambiguously demonstrate the detection of nuclear spins from the scanning sample we recorded NMR spectra while engaging or retracting a fluorinated tip on the NV center (Fig. 2a), thus modulating the presence or absence of $^{19}$F fluorine nuclei in the detection volume.

Spectra were acquired by dynamical decoupling noise spectroscopy established in earlier work[17,22] (Fig. 2b). Briefly, this scheme measures the power spectral density of magnetic field fluctuations $S_B(\omega)$, mapping the result to the $S_z$ projection of the NV center's ground state spin, which can be directly measured as fluorescence intensity of the center. Technically, this is accomplished by a protocol of microwave manipulations on the $|m_S = 0\rangle$ and $|m_S = 1\rangle$ levels of the center's spin ground state. It consists of a Ramsey interferometry sequence formed by a leading and trailing $\frac{\pi}{2}$ pulse, which sensitizes the center to magnetic fields, and an intermediate train of 80-200 $\pi$ pulses equidistantly spaced at a fixed delay $\tau$, which serve as a "quantum lock-in detection[23]" to spectrally enhance sensitivity to field fluctuations at the frequency $\omega = \pi/\tau$. The entire power spectral density $S_B(\omega)$ is sampled by repeating the experiment for varying $\tau$. NMR signals of surrounding nuclei manifest themselves as peaks in this magnetic noise spectrum.

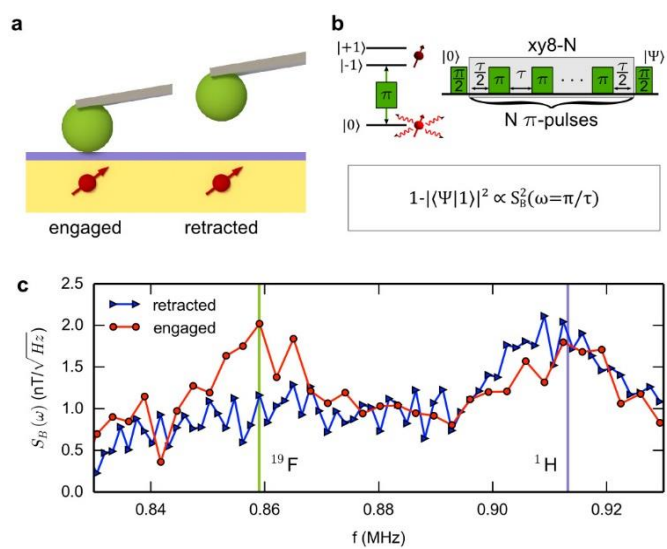

Fig 2- **Measurement scheme for Nuclear Magnetic Resonance spectroscopy. a**, Illustration of $^{19}$F signal switching by engaging and retracting the sample. **b**, Upper left panel: NV center ground state spin triplet, featuring optical spin readout ($m_s$=0 bright, $m_s$=±1 dark). Upper right panel: the power spectral density $S_B(\omega)$ of magnetic field fluctuations is measured by an XY8 microwave pulse sequence acting on the NV center's ground state spin. **c**, Experimental NMR spectra for the engaged and retracted sample. Sample nuclei are visible as a switchable peak of magnetic noise at the $^{19}$F Larmor frequency.

Spectra in presence or absence of the sample are displayed in Fig. 2c. Retracting the sample, we observe a spectrum with a single peak at the Larmor frequency of protons ($^1$H nuclei), which we attribute to a layer of proton-rich organic adsorbents on the diamond surface. Engaging the sample, a second peak rises at the Larmor frequency of Fluorine ($^{19}$F), proving detection of nuclei from the sample. The detection was fully reversible, with the peak reliably vanishing upon retraction of the sample.

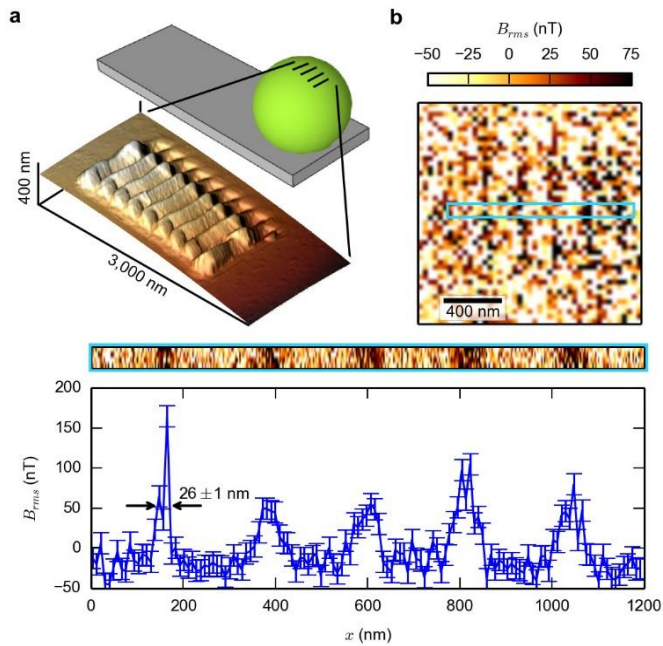

Fig 3- **Magnetic-Resonance Imaging of a nanoscale structure. a**, Sample used in the experiment. A calibration grating is engraved into a sample of $^{19}$F-rich Teflon®AF (green sphere) by a separate AFM. Lower panel: topography of the sample as obtained by AFM. **b**, $^{19}$F MRI image of the sample. **c**, Upper panel: Line-scan across the grating, same color bar as in **b**. Lower panel: Profile gained by 4x3 binning of the line-scan.

We were able to perform magnetic resonance imaging by recording the strength of the fluorine NMR peak $S_B(\omega_{19F})$ as a function of sample position (Fig. 3). Precisely, we employed a structure as shown in Fig. 3a, consisting of a calibration grating fabricated into a fluorinated sample (Teflon®AF) by nano-indentation in a separate atomic force microscope. A scanning-probe MRI image of this structure reliably reveals every single line of the grating (Fig. 3b+c) and faithfully reproduces both its width (710 nm) and spacing (200 nm). To benchmark the spatial resolution, we conducted a high-resolution line-scan across this pattern (Fig. 3b blue rectangle, Fig. 3c). The smallest resolved feature (leftmost peak) has a width of (26±1) nm (FWHM). This value is close to the ultimate resolution achievable by our method on a point-like sample, which should be approximately 10 nm, limited by the size of the detection volume and roughly equal to the distance between the NV sensor and the sample. To our best knowledge, Fig. 3c represents the smallest structure ever resolved by room temperature MRI.

As an extension of NMR, our method is able to distinguish different chemical species and to recover three-dimensional images of subsurface features. We were able to demonstrate these capabilities by simultaneously recording the strength of both $^{19}$F and $^{1}$H signals across a smooth unstructured Teflon®AF sample that was covered with a film of microscopy immersion oil to enhance visibility of the $^{1}$H adsorbate layer (Fig. 4).

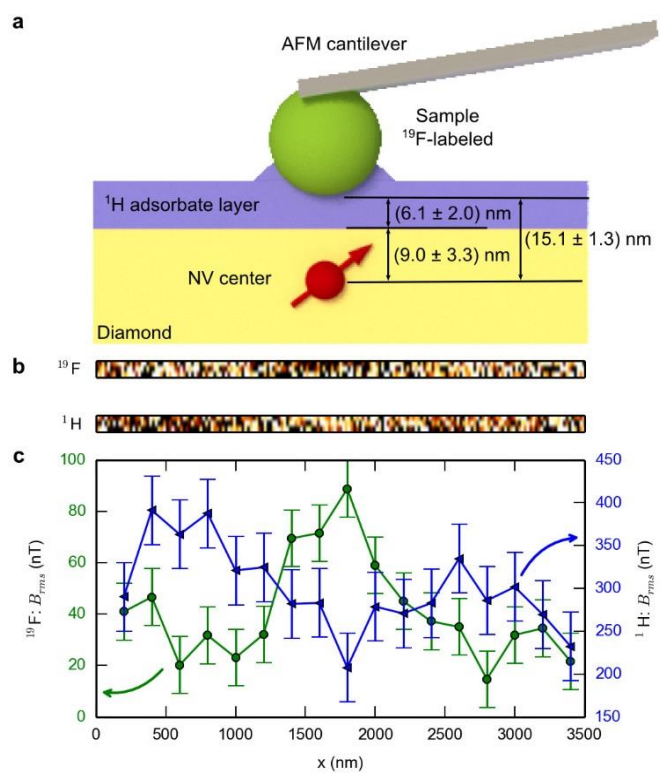

Fig 4 - **3D-imaging of a multilayer using chemical contrast a**, Schematics of the measurement setup. A scanning $^{19}$F-rich Teflon®AF sample locally compresses the $^{1}$H adsorbate layer. Distance values are obtained from quantitative analysis of MRI data **b**+**c**. **b**,

Line-scans displaying $^{19}$F and $^{1}$H signal strength.. **c,** Profile of line-scans in **b**, resulting from 3x10 binning with different y-axes for $^{19}$F (left scale) and $^{1}$H (right scale).

The resulting data is presented in Fig. 4b+c. Most prominently, we observe a drop in the $^{1}$H signal in the center region of the scan, concomitant with the increase of the $^{19}$F signal. We attribute this effect to compression of the proton-rich adsorbate layer under the contact area of the Teflon®AF sample. Closer inspection of the data reveals additional maxima of the $^{1}$H signal left and right of the contact area. We speculate that these features stem from the adsorbate layer forming a meniscus around the Teflon®AF tip as indicated in Fig. 4a, thus locally increasing the proton signal. We converted the data into a quantitative estimate of the NV-sample distance as well as the depth of the $^{1}$H adlayer by comparing it to a quantitative simulation of the signal strength under a $^{1}$H-$^{19}$F double layer (methods). The resulting values are displayed in Fig. 4a. This ability - to quantitatively recover depth information of a multilayer system - is a simple form of three-dimensional imaging, an inherent strength of MRI.

In summary, we have demonstrated scanning probe imaging of chemical species in a label-free approach achieving quantitative three-dimensional imaging with nanoscale resolution under ambient conditions. In future, two attractive extensions appear realistic. First, improved readout techniques for the NV sensor promise to accelerate acquisition by three orders of magnitude down to a speed of 20 milliseconds/pixel by using advanced spin readout techniques [24] or nuclear quantum memories[25]. Second, exploiting advanced NMR techniques will provide a much greater variety of contrast mechanisms than simple isotope-labeling used in this study. In particular, J-couplings or chemical shifts will allow imaging of specific chemical bonds and substances, while Zeeman shifts in a magnetic field gradient could provide depth information with atomic resolution[26]. We therefore expect the technique to find applications in various fields of research. In e.g. nanotribology, the ability to image interfacial layers in a three-dimensional and, possibly, structurally sensitive fashion will enable direct access to the structure and dynamics of abrasive samples and their lubricating films, which so far had to be inferred from indirect measurements of sliding force or topography[27]. In materials science, a versatile method for nuclear magnetic resonance imaging with nanoscale resolution could find use in fields like high-temperature superconductivity, where bulk NMR has already provided decisive insights[28] and where objects of interest are known to vary on small length scales, set for instance by the nanometer small coherence length[29]. Finally, the method should even be applicable to biological samples, such as microtome slices of the interior of a cell[30]. Ultimately, this direction of research could lead to in situ structure determination of single molecules by atomic-resolution magnetic resonance imaging, solving an outstanding quest of structural biology.

## Methods summary

**Experimental setup.** The measurements were conducted on an attocube™ "Combined Confocal and Scanning Force Microscope" (attoCSFM), allowing both optical readout via a confocal microscope and nanoscale manipulation via AFM. The NV center's spin is manipulated with microwave pulses delivered by a stripline. Optical readout and AFM are synchronized for pixel-wise measurements.

**Data acquisition.** Raster scanning the sample can induce artefacts for example through quenching and optical near-field effects of the sample that vary for different pixels. Extensive care was taken during data acquisition and processing to avoid these artefacts by comparison of eight datasets at each pixel and thus extract the NMR signal. This process is described in full detail in the methods section.

**Quantitative depth analysis.** Monte-Carlo-Simulations were used to extract quantitative information from the experiments. For the simulation, the sample was modelled by explicitly computing the field of randomly placed and oriented $^{19}$F nuclei inside an 80 nm cube above the NV center with the density calculated for Teflon®AF. The distance between fluorinated layer and NV center was then determined by comparison with the experimental result. The thickness of the proton-rich adsorbate layer between Teflon®AF sample and diamond was determined in the same way by comparing the experimental signal to a numerical simulation of a $^{1}$H-$^{19}$F double layer. This step also allows estimation of the NV center's implantation depth.

**Sample preparation.** DuPont™'s Teflon®AF 1600 was used as a fluorine-rich sample and fabricated to the end of tipless cantilevers, forming half-spherical single droplets. The calibration samples were created by a subsequent structuring process via nano-indentation. For the chemical contrast experiments (Fig. 4) the proton signal was enhanced by an additional coverage of the sample with a thin layer of immersion oil.

**End notes**

**Acknowledgement.** We acknowledge support from EU (via ERC grant SQUTEC and integrated projects Diadems and SIQS), Darpa (Quasar), the DFG via research group 1493 and SFB/TR21 and contract research of the Baden-Württemberg foundation.

**Author Contributions.** F.R. and J.W. conceived the idea and supervised the project. T.H. conducted the experiments and analyzed the data. D.S.-L. and T.H. prepared the samples. T.H., F.R. and J.W. wrote the manuscript.

# Figure legends

**Figure 1 | Experimental configuration**, A single NV center embedded few nm below the diamond surface (red spin) detects nuclear spins in a nanoscale detection volume above the diamond surface (red half-sphere). Imaging is performed by scanning a sample of NMR-active nuclear spins through this detection volume. In all experiments, a layer of proton rich organic adsorbates is present on the diamond surface, as it is unavoidable under ambient conditions.

**Figure 2 | Measurement scheme for Nuclear Magnetic Resonance spectroscopy. a**, Illustration of $^{19}$F signal switching by engaging and retracting the sample. **b**, Upper left panel: NV center ground state spin triplet, featuring optical spin readout ($m_s$=0 bright, $m_s$=±1 dark). Upper right panel: the power spectral density $S_B(\omega)$ of magnetic field fluctuations is measured by an XY8 microwave pulse sequence acting on the NV center's ground state spin. **c**, Experimental NMR spectra for the engaged and retracted sample. Sample nuclei are visible as a switchable peak of magnetic noise at the $^{19}$F Larmor frequency.

**Figure 3 | Magnetic-Resonance Imaging of a nanoscale structure. a**, Sample used in the experiment. A calibration grating is engraved into a sample of $^{19}$F-rich Teflon®AF (green sphere) by a separate AFM. Lower panel: topography of the sample as obtained by AFM. **b**, $^{19}$F MRI image of the sample. **c**, Upper panel: Line-scan across the grating, same color bar as in **b**. Lower panel: Profile gained by 4x3 binning of the line-scan.

**Figure 4 | 3D-imaging of a multilayer using chemical contrast a**, Schematics of the measurement setup. A scanning $^{19}$F-rich Teflon®AF sample locally compresses the $^1$H adsorbate layer. Distance values are obtained from quantitative analysis of MRI data **b**+**c**. **b**, Line-scans displaying $^{19}$F and $^1$H signal strength.. **c,** Profile of line-scans in **b**, resulting from 3x10 binning with different y-axes for $^{19}$F (left scale) and $^1$H (right scale).

# Methods

**NMR measurement scheme.** NMR spectroscopy was performed by dynamical decoupling noise spectroscopy using the XY8-N decoupling sequence [17] (Extended Data Fig. 1).

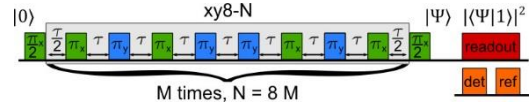

**Extended Data Fig 1 | XY8 sequence**, full representation of the applied pulse sequence. The pi-pulses are denoted with x and y, corresponding to a 90° phase-shift between the two, ensuring better error correction. During the readout laser pulse the photons are counted in the detection (det) and the reference (ref) window.

Here, an initial and final $\pi/2$ microwave pulse act as a Ramsey interferometer by creating a coherent superposition of the NV center's two spin states $(|0\rangle + e^{i\phi}|1\rangle)/\sqrt{2}$, whose phase is sensitive to fluctuating magnetic fields and thus to random spin fluctuations leading to a nonzero random phase $\Delta\phi$ with variance $\langle\Delta\phi^2\rangle$. Averaged over many repetitions this leads to a decay in the readout contrast $C$:

$$C = 2|\langle 1|\Psi\rangle|^2 - 1 = e^{-\langle\Delta\phi^2\rangle/2}$$

The train of $\pi$ pulses in between the initial and final $\pi/2$ pulses acts as a "quantum lock-in detection"[23], increasing and narrowing the sensitivity of $\langle\Delta\phi^2\rangle$ to a certain frequency band of magnetic noise defined by the wait-time $\tau$ and the number of $\pi$ pulses $N$ by introducing the filter function $S_g(\nu_n)$ with $\nu_n = n/N\tau$ in the following fashion:

$$\langle\Delta\phi^2\rangle = \gamma^2 \sum_{n=-\infty}^{\infty} S_g(\nu_n) S_B(\nu_n)$$

with $\gamma$ the NV spin's gyromagnetic ratio and $S_B(\nu_n)$ being the power spectral density (PSD) of the magnetic field fluctuations.

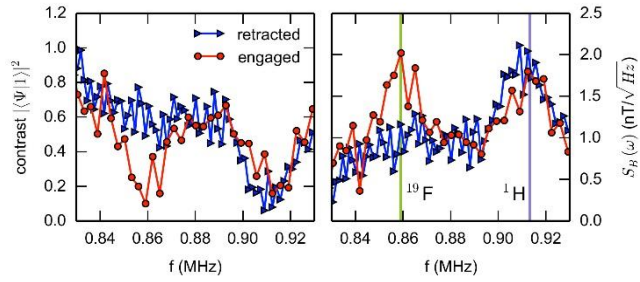

**Extended Data Fig 2 | Echo decay and reconstructed PSD** (same spectra as in Fig. 2c main text). The left panel shows the measured contrast. A clear decay is always discernable for the proton line, for the fluorine line only when the sample is engaged. Right panel: reconstructed PSD from left panel.

Every peak in the PSD is observed as an enhancement in the echo decay from which the PSD is then reconstructed as shown in Extended Data Fig. 2. The PSD was also modeled as a sum of Gaussian functions at the corresponding Larmor frequencies, convoluted with the sensitivity function and then fitted to the observed spectra. This allows the determination of the line-widths of the different peaks, which are considered constant for the used NV center.

**Experimental setup.** Our instrument is the attocube™ "Combined Confocal and Scanning Force Microscope" (attoCSFM). A device specially developed for highest stability and usability. In our experimental configuration (Extended Data Fig. 3) shallow implanted single NV centers (2.5 keV, $^{15}N^+$ ions) in a type IIa diamond membrane (thickness 30 µm) are excited and read out from below via a high NA oil-objective. Microwave pulses are delivered by a stripline fabricated on a glass cover slide below the diamond membrane. The AFM cantilever carrying the sample, which is read out interferometrically, can be approached to the diamond surface from above. A permanent magnet, attached to a 3D positioner stage, is used to apply an external field along the NV-axis.

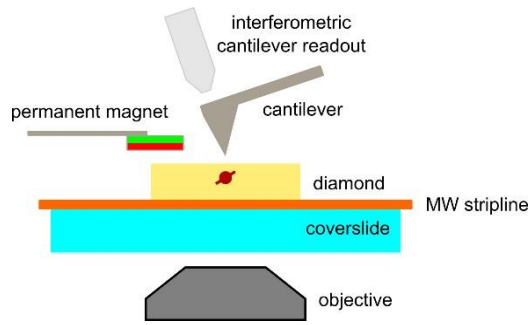

**Extended Data Figure 3 | Experimental setup.** Schematics showing the diamond containing the shallow NV's on top of the MW stripline fabricated onto a coverslide. Optical readout is done from below via a high NA objective, the AFM cantilever with interferometric readout is approached from above, a permanent magnet can be positioned relative to the NV via a 3D positioning system.

The optical readout of the spin state of the NV center is synchronized to the motion of the AFM tip, thus making pixel-wise measurements possible.

**Data Acquisition for scanning probe MRI images.** To record the strength of the $^{19}$F NMR signal as a function of tip position (MRI images Fig. 3+4 main text), we perform an XY8 measurement at every pixel of the AFM scan. This quantum protocol (Extended Data Fig. 1) maps $S_B(\omega_F)$, the strength of magnetic noise at the $^{19}$F Larmor frequency, to the spin projection $\langle S_z \rangle = |\langle \psi | 1 \rangle|^2$ of the NV center, which translates into a change of the center's fluorescence intensity.

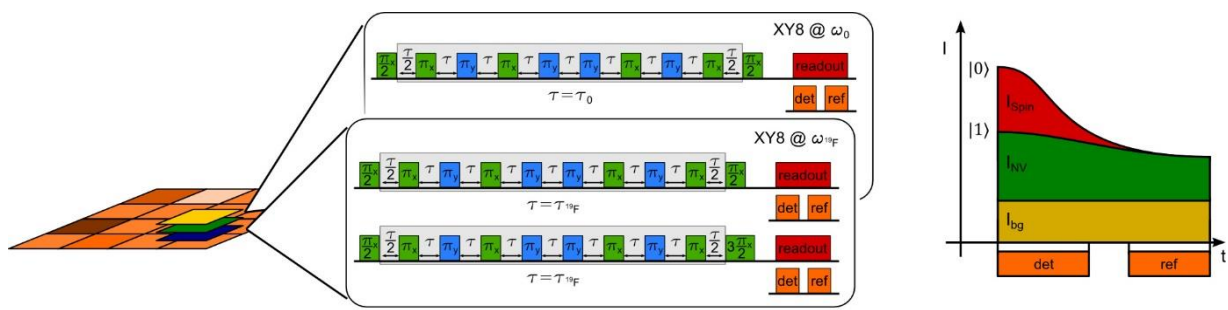

**Extended Data Figure 4 | Schematics of dataset acquisition.** At every pixel XY8 sequences are recorded at two different inter-pulse delays $\tau$, each consisting of two measurements (labeled

$\pi/2$ and $3\pi/2$) differing by a final $\pi$ pulse. Additionally, the photon flux of the readout pulse is divided into a detection and a reference window, finally adding up to eight datasets.

In a simple scheme, $|\langle\psi|1\rangle|^2$ could be inferred directly from the fluorescence intensity emitted by the center upon a laser readout pulse following the XY8 microwave manipulation. This simple scheme, however, is prone to artefacts, since it cannot distinguish a spin-dependent fluorescence change from numerous additional sources of luminescence, such as background fluorescence from the sample.

To avoid these artefacts, we recover $|\langle\psi|1\rangle|^2$ from a careful comparison of eight datasets (Extended Data Fig. 4), generated by performing a series of four XY8 measurements with varying parameters at every pixel of the scan, and recording fluorescence within two windows of the laser readout pulse (labeled "det" and "ref" in Extended Data Fig. 1+4). This procedure is equivalent to the acquisition of a single-point NMR spectrum at every pixel and suppresses all conceivable sources of artefacts (see below).

Precisely, the four XY8 measurements consist of two series of measurements with different values of inter-pulse delay $\tau$, sampling spin noise at the $^{19}$F Larmor frequency $\omega_{19F}$ and a reference frequency $\omega_0$ far from any NMR transition. Each series consists of a set of two measurements, labeled $\pi/2$ and $3\pi/2$ in the following, which differ by a final $\pi$ pulse such as to convert NMR intensity into a fluorescence increase or drop, respectively (Extended Data Fig. 2).

The photon flux of the readout pulse at the end of each pulse sequence is divided into two windows, the detection window (where NV fluorescence is partly spin dependent) and the so-called reference window (where NV fluorescence is spin independent) (Extended Data Fig. 1+4), resulting in the intensities $I_{det}$ and $I_{ref}$, respectively.

We will now show that the NV center's spin contrast $C = 1 - 2|\langle\psi|1\rangle|^2$ at the end of the XY8 measurement sequence can be reliably recovered from these datasets by the relation

$$C = \frac{\left[(I_{det} - I_{ref})_{3\pi/2} - (I_{det} - I_{ref})_{\pi/2}\right]_{f_{19F}}}{\left[(I_{det} - I_{ref})_{3\pi/2} - (I_{det} - I_{ref})_{\pi/2}\right]_{f_0}} \quad (1)$$

To this end, we consider a maximally pessimistic model of NV fluorescence intensity, which includes background luminescence, amplification or quenching of NV luminescence by near-field effects from the sample as well as discharging between the charge states NV⁻ and NV⁰ [31]. We describe all these effects on the total detected intensity $I_{tot}$ by the ansatz

$$I_{tot} = I_{bg} + \alpha(I_{NV} + \beta \cdot I_{spin})$$

which explicitly accounts for the following effects (schematically shown in Extended Data Fig. 4, right panel):

$I_{bg}$: background luminescence from the sample. This luminescence is constant in time.

$I_{NV}$: spin-independent fluorescence intensity of the NV center (e.g. background from a spin-inactive NV⁰ population). Over the laser detection pulse, this contribution varies in an unknown manner

$I_{spin}$: spin-dependent fluorescence intensity of the NV center. Only this term depends on the spin projection $C = 1 - 2|\langle\psi|1\rangle|^2$ and hence represents "useful" signal. Over the laser detection pulse, its contribution varies according to the relations

$I_{spin,det} = C \cdot I_{spin,0}$;

$I_{spin,ref} = 0$;

$\alpha$: amplification or quenching of all NV fluorescence by optical near-field effects of the sample.

$\beta$: discharging between the NV⁻ and NV⁰ charge states, altering the ratio of spin-dependent and spin-independent signal.

Importantly, all these quantities vary spatially, resulting in different contributions at every pixel of the scan.

With this model, proving equation (1) is straightforward. Yet, it is instructive to consider its constituents in detail.

Subtraction of reference and detection time windows yields:

$$I_{det} - I_{ref} = \alpha \cdot \Delta I_{NV} + \alpha \cdot \beta \cdot I_{spin,0} \cdot C$$

cancelling the time-independent contribution $I_{bg}$. $\Delta$ denotes the difference of luminescence between detection and reference window.

To remove $\alpha \cdot \Delta I_{NV}$, a potential artefact of NV⁰ luminescence, the sequence is measured again, replacing the last $\pi/2$ pulse by a $3\pi/2$ pulse, effectively inverting the spin-dependent part of the result ($C \to -C$). The difference of both sequences results in

$$\left(I_{det} - I_{ref}\right)_{3\pi/2} - \left(I_{det} - I_{ref}\right)_{\pi/2} = 2 \cdot \alpha \cdot \beta \cdot I_{spin,0} \cdot C$$

which is still compromised by the spatially varying prefactors $\alpha$ and $\beta$ (Extended Data Fig. 3, third line). These, however, can be eliminated by the reference measurement at a different frequency $f_0$, where no signal is expected ($C(f_0) = 1$), finally resulting in:

$$C = \frac{\left[\left(I_{det} - I_{ref}\right)_{3\pi/2} - \left(I_{det} - I_{ref}\right)_{\pi/2}\right]_{f_{19F}}}{\left[\left(I_{det} - I_{ref}\right)_{3\pi/2} - \left(I_{det} - I_{ref}\right)_{\pi/2}\right]_{f_0}}$$

The effect of this procedure is illustrated in Extended Data Fig. 5, which displays varying levels of correction for the line-scan of Fig. 3, main text. Clearly, a simple recording of NV fluorescence (Extended Data Fig. 5a) is insufficient to unambiguously prove presence of the

$^{19}$F signal, since near-field effects and background luminescence mask the NMR signal. Only a careful comparison of all eight datasets by means of eq. (1) provides a clear MRI image (Extended Data Fig. 5e).

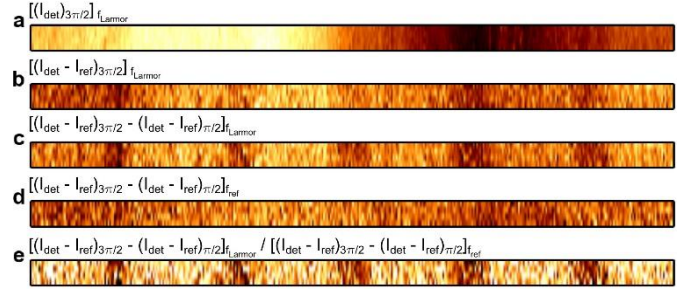

**Extended Data Figure 5 | Data evaluation scheme**, demonstrated on the line-scan of Fig. 3, main text. **a**, NV fluorescence from detection window, strong near-field influence of the tip visible. **b**, subtraction of detection and reference window. **c**, combination of $\pi/2$ and $3\pi/2$ pulse-sequence. **d**, corresponding contrast measured at reference frequency. **e**, division of **c** and **d**, yielding final and artefact free result.

With this result the $B_{rms}$ values for the images in Fig. 3+4, main text, were computed as follows. The PSD $S_B(\nu)$ was again modelled as a Gaussian function $S_B(\nu) = Ae^{-\frac{(\nu-\nu_0)^2}{2c^2}}$, with $\nu_0$ being the Larmor frequency of the nucleus of interest, and fitted to the measured contrast $C$ at every pixel independently by adapting the amplitude $A$. The line-width $c$ of the Gaussian function was considered constant for the particular NV center in use, as described above (see Methods, Measurement scheme), and was taken from the acquired full spectra. The root-mean-square value of the magnetic field noise $B_{rms}$ is then given by the following integral, which is easily evaluated using the known relations for Gaussian functions:

$$B_{rms} = \sqrt{\int_{-\infty}^{\infty} (S_B(\nu))^2 d\nu} = A \cdot \sqrt{c} \cdot \sqrt[4]{\pi}$$

**Quantitative depth analysis.** Quantitative information could be extracted through comparison of the experimental results with numerical simulations by adapting the simulations conducted in previous work [17]. The fluorinated sample was modelled as depicted in Extended Data Fig. 6 by explicitly computing the field of randomly placed and oriented $^{19}F$ nuclei inside an 80 nm cube above the NV center with a density of $^{19}F$ nuclei calculated for Teflon®AF of $\rho_F = 3.7 \times 10^{28}/m^3$. The distance between fluorinated layer and NV center was then determined by comparison with the experimental results (Fig. 4, main text). The proton-rich adsorbate layer was modelled in the same way with a proton density of $\rho_H = 5.0 \times 10^{28}/m^3$, but with variable thickness of the layer (Extended Data Fig. 6). The thickness could again be determined by comparison with experimental results, which allowed the calculation of the implantation depth of the NV center (Fig. 4, main text).

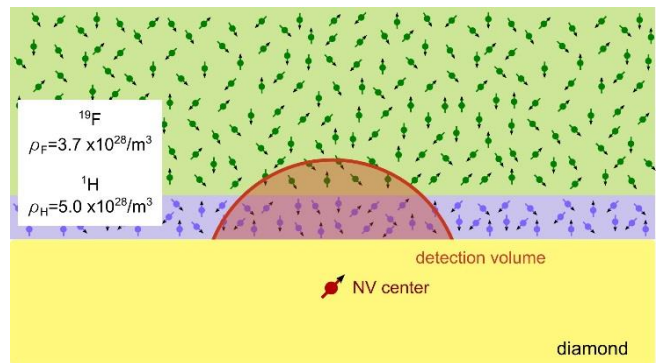

**Extended Data Figure 6 | Schematics for Monte Carlo Simulation**, $^{19}F$ nuclides were randomly placed and oriented in a volume above the NV center. The proton rich adsorbates were modeled as a layer of $^{1}H$ nuclides between diamond and the fluorinated volume. The simulation was done in all three dimensions.

**Sample Preparation.** We used DuPont[TM]'s Teflon®AF, namely Teflon®AF 1600, an amorphous fluoroplastic (full name Poly[4,5-difluoro-2,2-bis(trifluoromethyl)-1,3-dioxole-co-tetrafluoroethylene]), as a fluorine-rich sample that contains no protons. It was procured from Sigma Aldrich and ships as a white powder, which was then solved in Perfluoromethylcyclohexan and further diluted in Perlfuorodecalin to form a viscose solution.

Using a micromanipulator we fabricated single droplets of the Teflon®AF solution to the end of tipless cantilevers. The diameters of these half-spherical shaped droplets were in the range of 10-15 µm. The cantilevers are from Budget Sensors, named "all-in-one" afm probes, tipless, or short: AIO-TL. Every chip has four cantilevers of different properties, we used cantilever B with a force constant of 2.7 N/m.

After application of the droplet, the sample was baked in a stepwise fashion. First it was held for 15 min at 160°C to make sure all the solvent had evaporated. In the second step the heat was ramped up to 300°C for 30 min. This temperature is well above the glass transition temperature of the Teflon®AF 1600 which is at 160°C and anneals the Teflon®AF, thus forming a smooth and relative hard surface, as confirmed by AFM scans.

Using a separate AFM, the Teflon®AF droplet was structured by nano-indentation to create calibration samples.

In the chemical contrast experiments the samples were additionally covered with a thin layer of immersion oil (Sigma Aldrich 10976) to increase the visibility of the proton signal.

# Extended data figures legends

**Extended Data Figure 1 | XY8 sequence**, full representation of the applied pulse sequence. The pi-pulses are denoted with x and y, corresponding to a 90° phase-shift between the two, ensuring better error correction. During the readout laser pulse the photons are counted in the detection (det) and the reference (ref) window.

**Extended Data Figure 2 | Echo decay and reconstructed PSD** (same spectra as in Fig. 2c main text). The left panel shows the measured contrast. A clear decay is always discernable for the proton line, for the fluorine line only when the sample is engaged. Right panel: reconstructed PSD from left panel.

**Extended Data Figure 3 | Experimental setup**. Schematics showing the diamond containing the shallow NV's on top of the MW stripline fabricated onto a coverslide. Optical readout is done from below via a high NA objective, the AFM cantilever with interferometric readout is approached from above, a permanent magnet can be positioned relative to the NV via a 3D positioning system.

**Extended Data Figure 4 | Schematics of dataset acquisition.** At every pixel XY8 sequences are recorded at two different inter-pulse delays $\tau$, each consisting of two measurements (labeled $\pi/2$ and $3\pi/2$) differing by a final $\pi$ pulse. Additionally, the photon flux of the readout pulse is divided into a detection and a reference window, finally adding up to eight datasets.

**Extended Data Figure 5 | Data evaluation scheme**, demonstrated on the line-scan of Fig. 3, main text. **a**, NV fluorescence from detection window, strong near-field influence of the tip visible. **b**, subtraction of detection and reference window. **c**, combination of $\pi/2$ and $3\pi/2$ pulse-sequence. **d**, corresponding contrast measured at reference frequency. **e**, division of **c** and **d**, yielding final and artefact free result.

**Extended Data Figure 6 | Schematics for Monte Carlo Simulation**, $^{19}$F nuclides were randomly placed and oriented in a volume above the NV center. The proton rich adsorbates were modeled as a layer of $^1$H nuclides between diamond and the fluorinated volume. The simulation was done in all three dimensions.